\documentclass[universe,article,submit,moreauthors,pdftex,10pt,a4paper]{mdpi-new}
\newcommand{\lsim}{\stackrel{\scriptstyle <}{\phantom{}_{\sim}}}
\newcommand{\gsim}{\stackrel{\scriptstyle >}{\phantom{}_{\sim}}}
\firstpage{1}
\articlenumber{x}
\doinum{10.3390/------}
\pubvolume{xx}
\pubyear{2017}
\copyrightyear{2017}
\externaleditor{Academic Editor: name}
\history{Received: date; Accepted: date; Published: date}

\nolinenumbers\Title{Evolution of quasiperiodic structures in non-ideal hydrodynamic description of phase transitions}

\Author{D. N. Voskresensky $^{1,2
}$
}

\address{%
$^{1}$ \quad Joint Institute for Nuclear Research, RU-141980 Dubna, Moscow region, Russia
\\
$^{2}$ \quad
National Research Nuclear  University (MEPhI), Kashirskoe shosse 31, 115409 Moscow, Russia}

\corres{Correspondence: d.voskresen@gmail.com}

 \abstract{Various  phase transitions could have taken place in the early Universe,  and may occur in the course of heavy-ion collisions and  supernova explosions, in  proto-neutron stars,  cold compact stars, and in the condensed matter at  terrestrial conditions.  Most generally, the dynamics of the density and temperature at first-  and second-order phase transitions can be described with the help of the equations of non-ideal hydrodynamics. In the given work some novel solutions are found describing the evolution of quasiperiodic structures that are formed in the course of the  phase transitions. Although this consideration is very general, particular examples of  quark-hadron and nuclear liquid-gas first-order phase transitions to the uniform $k_0=0$ state and  pion condensate second-order phase transition to non-uniform  $k_0\neq 0$ state in dense baryon matter are considered.
}


\begin{document}

\section{Introduction}\nolinenumbers
Cosmological observations of the two last
decades \cite{Planck} supplied us with some extraordinary results and
puzzles, in particular with the fact that the Universe undergoes
an accelerated expansion and that only 5$\%$ of its mass is in baryons, 26$\%$ is in a dark matter and remaining part is in a dark energy.
It is commonly believed that at least two cosmic phase transitions have occurred in the early Universe, the electro-weak and the
QCD phase transitions \cite{Linde,Witten:1984rs}. The Standard Model   of particle physics predicts that after the inflation the hot expanding Universe was   filled with deconfined quarks, in the state of quark-gluon plasma \cite{Rafelski:2013obw}. This view on the early Universe is supported by  simulations done in various cosmological  and  relativistic heavy ion collision models \cite{Busza:2018rrf,Jacak}  and by the lattice calculations \cite{Fodor}.
The quark-gluon plasma in baryon-poor matter persists down to a temperature $T\simeq  160$ MeV. However the Standard Model  does not account for the
presence of the dark matter, with which additional cosmic phase transitions may be associated during cooling of the expanding Universe  to its present temperature $T\simeq 2.7$K.

Another  piece of important information about strongly interacting matter can be extracted from neutrino and photon radiation of compact stars formed in supernova events \cite{ST83,Migdal:1990vm} and from analysis of gravitational waves in gamma ray bursts. A strong phase transition may result
in a second neutrino burst if it occurred during supernova explosion and a  hot neutron star formation  or right after \cite{Haubold:1988uu,Migdal:1990vm}. It might be associated with a significant delay of the heat transport to the neutron-star surface, if the system is close to the pion-condensate phase transition. Recently new arguments have been expressed for that indeed two neutrino bursts were measured during 1987A explosion, one delayed respectively other by 4.7h, cf. \cite{Galeotti:2016uum}.  The second burst and a blowing of some amount of matter could be then related to the phase transition of the neutron star to the pion condensate state. In old neutron stars the first-order phase transition, if occurred,  could result in a strong star-quake  \cite{Migdal:1990vm,Prakash:1996xs}. Detection  of  merging  compact stars in  the gravitational wave spectra \cite{Abbott:2017xzu} and detection of massive compact stars \cite{Demorest:2010bx,Fonseca:2016tux,Antoniadis:2013pzd,Cromartie:2019kug} provide  constraints  on  the  equation   of state of strongly interacting dense matter and strong phase transitions in it.

Experimental study of the  ultrarelativistic  heavy-ion collisions helps to  simulate at the terrestrial conditions the processes have occurred  in the very early Universe, in supernova explosions and in gamma ray bursts.
Experimental data and lattice calculations \cite{Fodor} indicate that   the hadron-quark  transition in heavy-ion collisions at RHIC and LHC collision energies is the crossover transition, cf.  \cite{Shuryak:2004cy,Romatschke:2007mq,Teaney:2001av}.   For lower collision energies relevant for NICA and FAIR facilities one expects to find signatures of the strong first-order quark-hadron phase transition \cite{Shuryak:2008eq}.
There are experimental evidences that in the very low-energetic heavy-ion collisions of approximately isospin-symmetrical nuclei there occurs the first-order nuclear liquid-gas phase transition  (for  temperatures $T\lsim 20$ MeV and baryon densities $n\lsim 0.7 n_0$, where $n_0$ is the nuclear saturation density) ~\cite{Ropke:1982ino,SVB,Chomaz:2003dz}.

In a many-component system a mechanical instability is accompanied by a chemical instability, see Ref.~\cite{Margueron:2002wk,Maslov:2019dep}. The inclusion of the Coulomb interaction, see Refs.~\cite{Ravenhall:1983uh,Maruyama:2005vb}, leads to a possibility of the pasta phase in the neutron star crusts for densities $0.3 n_0\lsim n\lsim 0.7 n_0$.
  For higher densities in dense neutron star interiors there may appear phase transitions to the pion \cite{Migdal:1978az,Migdal:1990vm}, kaon \cite{Glendenning:2001pe,Maruyama:2005tb}, charged rho \cite{Voskresensky:1997ub} condensate  states and to the quark matter \cite{Glendenning:2001pe,Glendenning:1992vb,Heiselberg:1992dx,Voskresensky:2002hu,Maslov:2018ghi}. The quark-hadron, pion, kaon and charged rho-meson
 condensate  phase transitions may occur during iso-entropical   falling of the baryon-rich matter in the supernova explosions \cite{ST83}, in proto-neutron stars and in cold compact stars, cf.   \cite{Migdal:1990vm}.   In some models these phase transitions are considered as  first order phase transitions leading to  mixed   phases in dense matter.  Formation of the pasta non-uniform phases is one of the possibilities \cite{Voskresensky:2002hu,Maruyama:2005tb,Maslov:2018ghi}.   Add here possibilities of the phase transitions between various superfluid \cite{Sedrakian:2018ydt,Kolomeitsev:2010pm} and  ferromagnetic-superfluid \cite{Voskresensky:2019zcp} phases in the cold neutron stars and in the color-superconducting hybrid compact stars \cite{Alford:2007xm},
 as well as  numerous possibilities of the phase transitions in the condensed matter physics at terrestrial conditions, like liquid-gas, liquid-glass, glass-metal transitions, etc.

 The liquid-gas phase transition, transition to the superfluid state in quantum liquids and many other transitions occur to the uniform state characterized by the wave number $k_0=0$.
 Other phase transitions, like the transitions in solids and liquid crystals,  are the transitions to the inhomogeneous states characterized by the non-zero wave-vectors, $\vec{k}_{0,i}\neq 0$, cf. \cite{Voskresensky:1984rd,Voskresensky:1993ux}. In glasses the order, characterized by $k_0\neq 0$ appears at rather short distances  but disappears at long distances \cite{Voskresensky:1993ux}. The phase transition to the  pion condensate state \cite{Migdal:1978az,Migdal:1990vm} possible in interiors of neutron stars may occur due to a strong $p$-wave pion-baryon attraction, which increases with  increase of the baryon density. Thereby the pion condensation occurs in non-uniform state, $k_0\neq 0$. Perhaps   the antikaon condensation in dense baryon matter also occurs to the non-uniform state, $k_0\neq 0$, cf. \cite{Kolomeitsev:2002pg}.

 Some of the mentioned phase transitions, as the transition of the normal matter to superfluid in metals and in $^4$He, are transitions of the second order \cite{Tilli}. Other mentioned above phase transitions, such as
 the liquid-gas phase transition, are transitions of the first order. Search for the critical endpoint separating the crossover and first-order quark-hadron transitions is one of the benchmarks for   the future experiments at NICA and FAIR.

 In early Universe, at the processes of the formation of compact stars in supernova explosions, collisions of compact binary stars, and in heavy-ion collisions one deals with a rapid thermalization of a strongly interacting quark-gluon matter and then the hadronic matter. These processes can be described within non-ideal hydrodynamics, where viscosity and thermal conductivity effects are of crucial importance. The dynamics of the phase transitions also can be considered within non-ideal hydrodynamics, cf. \cite{Voskresensky:1993ux,Skokov:2008zp,Skokov:2009yu,Skokov:2010dd}.

Below some novel solutions will be found describing   evolution of  periodic structures at the second-order phase transitions to the  non-uniform  state with the wave number  $k_0\neq 0$, and quasiperiodic time-dependent structures appearing  in the course   of the  spinodal instabilities at the first order phase transitions to the uniform  state with  $k_0=0$ and in the dynamics of  the second-order phase transitions occurring in the uniform state. Although consideration is very general, as specific examples,  quark-hadron and nuclear liquid-gas first-order phase transitions and  the pion condensation second-order transition will be considered.

The presentation is organized as follows. In Sect. \ref{Waals} main features of the van der Waals-like equation of state are reminded. In Sect. \ref{hydro} hydrodynamical description  is formulated  for the description of the first- and second-order phase transitions to the uniform $k_0=0$ state and for the second-order phase transition to the nonuniform,  $k_0\neq 0$, state in assumption of a small overcriticality. Dynamics of seeds at the first-order phase transition from a metastable  to the stable state is considered in Sect. \ref{metastable}.
Dynamics of fluctuations in unstable region is studied in \ref{unstable}. Some novel solutions describing time evolution of quasiperiodic and periodic structures are found. Sect. \ref{Conclusion} contains some concluding remarks.

\section{van der Waals-like equation of state for description of  first-order phase transitions}\label{Waals}

\begin{figure}\centering
\includegraphics[width=5.8cm,clip]{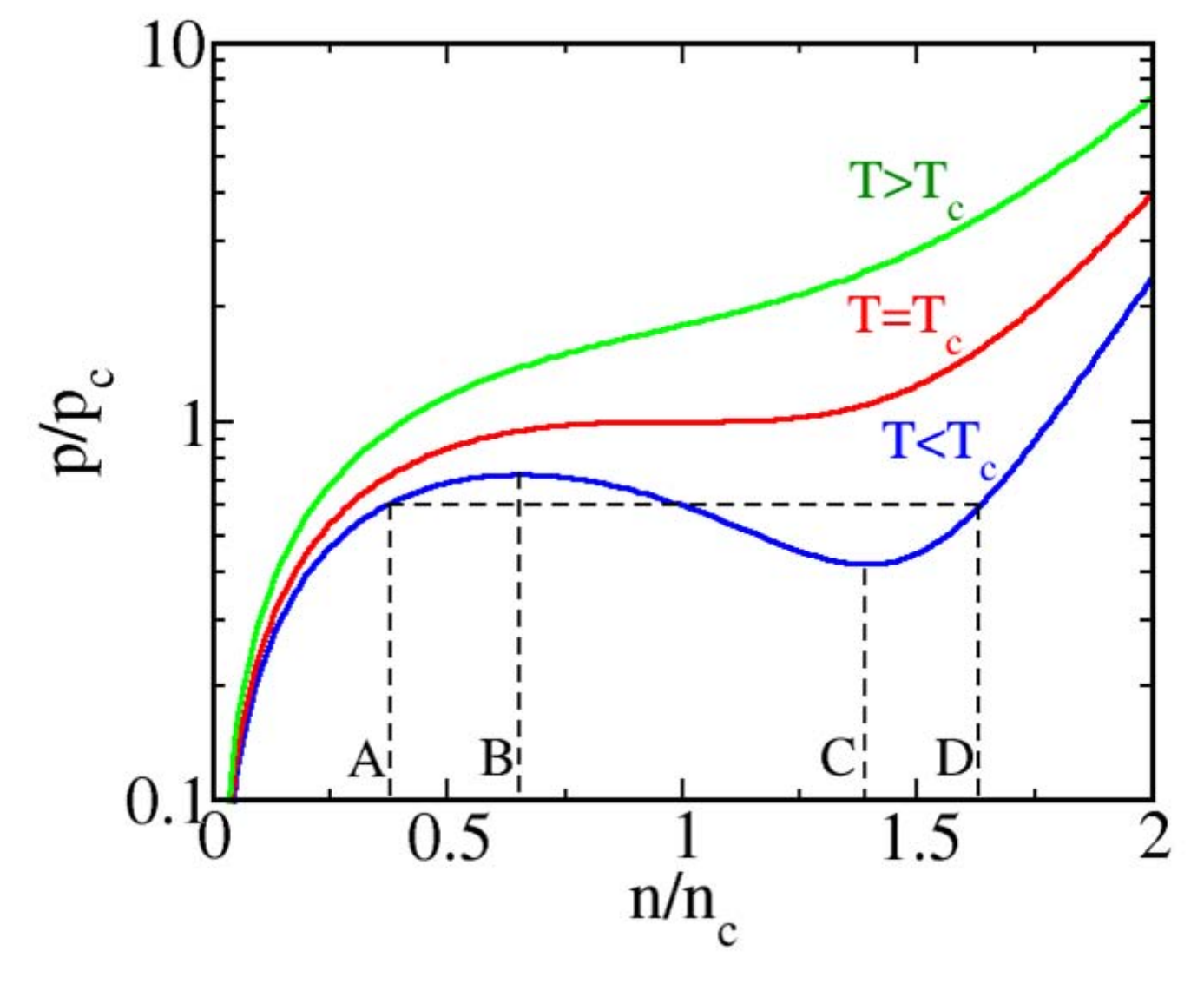}
\caption{Schematic pressure isotherms as  functions  of the number density $n$ at a liquid-gas-like phase transition. $P_{c}$, $n_{c}$ and $T_c$ are the pressure, density and  temperature at the critical point.
 }\label{VanderWaals}
\end{figure}

  The dynamical trajectories of the expanding baryon-rich matter in the  heavy-ion collisions and of the falling matter in supernova explosions till a phase transition did not occur can be  characterized by approximately conserved entropy, whereas the volume $V$ and the temperature $T$ are changed with the time. In the simplest  case of  one-component matter, e.g. the baryon matter,  the pressure -- baryon number density isotherms $P(n)|_T$ describing the liquid-like (with a higher density) or gas-like (with a smaller  density) states  demonstrate a monotonous behavior for  the values of the temperature $T$ above the critical  temperature of the first-order liquid-gas type phase transition. However  $P(n)|_T$ isotherms acquire a convex-concave form for $T$ below the critical temperature  \cite{Voskresensky:1993ux}, see Fig. \ref{VanderWaals}. The horizontal dashed line connecting points A and D shows the Maxwell construction (MC) describing thermal equilibrium of phases. At equilibrium the baryon chemical potentials are $\mu_A=\mu_D$. The interval AB corresponds to a metastable supercooled vapor (SV) and the interval CD relates to a  metastable overheated liquid (OL). The interval BC shows  unstable spinodal region.

\begin{figure}[h]
\centering
\includegraphics[width=0.45\textwidth]{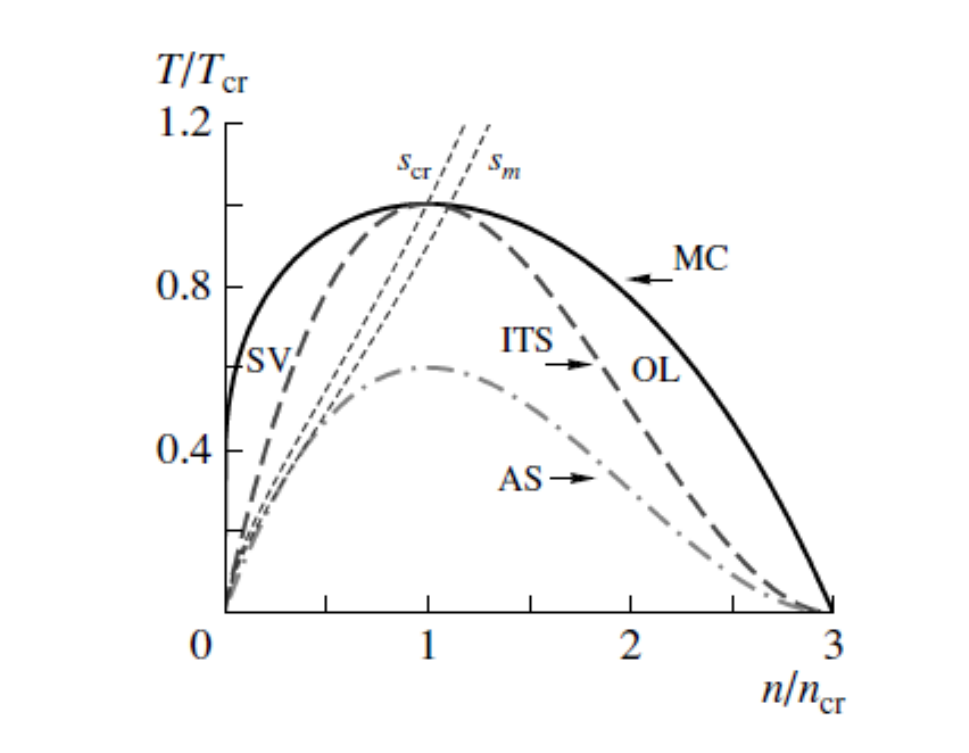}
\caption{ The phase diagram of the van der Waals equation of state on
$T(n)$-plane. The bold solid,  dashed and dash-dotted curves show boundaries of the MC, the spinodal
region at $T=$const and $\tilde{s}=$const, respectively. The short
dashed lines show two adiabatic trajectories of the system evolution: the curve labeled
 $s_{cr}$ passes through the critical point; $s_m$, through the maximum pressure point $P(n_{P,max})$ on the
 $P(n)$ plane. Figure is adopted  from \cite{Skokov:2010dd}.
 }
\label{spinT}
\end{figure}

 Adiabatic trajectories, short dashed lines $s_{cr}$ and $s_m$, where $\tilde{s}\equiv s/n \simeq$ const, $s$ is the
entropy density, are
shown in Fig. \ref{spinT} on the plot of $T/T_{cr} = f (n/n_{cr})$.
The upper convex curve (MC, bold solid line) demonstrates the boundary of the MC, the
bold dashed line, ITS, shows the boundary of the isothermal spinodal  region and the  bold dash-dotted
curve AS, indicates the boundary of the adiabatic spinodal region.  At the ITS line $u_T^2=(\partial P/\partial \rho)_T=0$ and at AS line, $u_{\tilde{s}}^2=(\partial P/\partial \rho)_{\tilde{s}}=0$, where $u_T$ and $u_{\tilde{s}}$ have the meaning of the isothermal and adiabatic sound velocities, respectively,  $\rho=m^* n$, $m^*$ is the baryon quasiparticle mass.
The supercooled vapor (SV) and the overheated liquid (OL) regions
are situated between the MC and the ITS curves, on the left and on the right, respectively.
For $\tilde{s}_{cr}>\tilde{s}>\tilde{s}_{\rm MC2}$, where
$\tilde{s}_{cr}$ is the  value  of the specific entropy $\tilde{s}$
at the critical point and the line
with $\tilde{s}_{\rm MC2}$ in the example shown in Fig. \ref{spinT} passes through the point $n/n_{cr}=3$
at $T=0$, the system traverses the OL state (the region OL in Fig. \ref{spinT}), the ITS region (below the ITS line) and the AS region (below the
AS line). For $\tilde{s}>\tilde{s}_{cr}$ the system trajectory
passes through the SV state (the region SV in Fig. \ref{spinT}) and  the
ITS region.
Note that in reality for the quark-hadron first-order phase transition the phase diagram looks a bit different, since then $T_{cr}$ increases with a decrease of the baryon density \cite{Steinheimer:2013gla,Steinheimer:2016bet}. However  this peculiarity   does not change a general analysis given here.

After the system enters the region of the fist-order phase transition of the liquid-gas type the approximation of  constant entropy fails and  the description of the dynamics of the  system  needs solution of non-ideal hydrodynamical equations \cite{Skokov:2008zp,Skokov:2009yu,Skokov:2010dd}.
Similarly, the description of the dynamics of the second-order phase transition needs solution of non-ideal hydrodynamical equations in the case, when the density and the temperature (or entropy) can be considered as appropriate order parameters.

\section{Hydrodynamical description  of first- and second-order phase transitions at small overcriticality}\label{hydro}
 Further assume that the dynamics of a second order phase transition and of a first-order phase transition  can be described  by the variables $n$ and $s$ (or $T$), cf.  \cite{Skokov:2008zp,Skokov:2009yu,Skokov:2010dd}.
Also, assume that the system is rather close to the critical point of the phase transition. Since all the processes   in the vicinity of the critical point  are slowed down, the velocity of a seed of a new phase prepared in the old phase, $\vec{u}$, is much less than the mean thermal velocity and one may use
equations of non-relativistic non-ideal hydrodynamics:
the Navier-Stokes equation, the continuity equation,
and equation for the heat
transport, even if one deals with violent heavy-ion collisions:
\begin{eqnarray}
\label{Navier} m^{*} n\left[ \partial_{t} {u}_i + (\vec{u}\nabla)
{u}_i \right] &=& -\nabla_i P   + \nabla_k \left[ \eta \left(
\nabla_k u_i + \nabla_i u_k -\frac{2}{\nu} \delta_{ik} \mbox{div}
\vec{u}   \right)   + \zeta \delta_{ik} \mbox{div} \vec{u}
\right], \\
\label{contin}
\partial_{t}n +\mbox{div} (n \vec{u})&=&0, \\
 \label{therm}   T\left[\frac{\partial
s}{\partial t} +\mbox{div}(s\vec{u} )\right]&=&\mbox{div}(\kappa
\nabla T) +\eta \left(\nabla_k u_i + \nabla_i u_k -\frac{2}{\nu}
\delta_{ik} \mbox{div} \vec{u} \right)^2 +\zeta (\mbox{div}
\vec{u})^2\,.
\end{eqnarray}
Here, as above, $n$ is the number density of the conserving charge, to be specific, the baryon
density, $m^{*}$ is the baryon quasiparticle mass, $P$ is the pressure. The quantities   $\eta$
and $\zeta$ are the shear and bulk viscosities, $\nu$ shows the
geometry of the seed under consideration  (droplets, rods, slabs),   $\kappa$ is the thermal conductivity.

All thermodynamical quantities can be expanded  near  a  reference point
$(n_{\rm r}, T_{\rm r})$,  which we assume to be close to the critical point but  still  outside the  fluctuation region, which we assume to be narrow.   This circumstance is important for the determination of the specific heat density $c_{V ,\rm r}$ and, m.b.,  transport coefficients, which may diverge in the critical point, whereas other quantities are  smooth functions of $n,T$ and calculating  them one can put $n_{\rm r}=n_{cr}, T_{\rm r}=T_{cr}$.

The Landau free energy counted from the value at $n_{\rm r}\simeq n_{cr}, T_{\rm r}\simeq T_{cr}$
in the variables $\delta n
=n-n_{cr}$, $\delta T=T-T_{cr}$,
$\delta (\delta F_L)/\delta(\delta n) = P - P_{f}+P_{\rm MC} $ can be presented as \cite{Skokov:2008zp,Skokov:2009yu,Skokov:2010dd}
\begin{eqnarray}
&&\delta F_{\rm L} = \int \frac{d^3 x}{n_{cr}}\left\{ \frac{cm^{*}[\nabla
(\delta n)] ^2}{2}+\frac{\lambda m^{*\,3} (\delta
n)^4}{4}-\frac{\lambda v^2 m^{*}(\delta n) ^2}{2}-\epsilon \delta n
\right\}+ \delta F_{\rm L} (k_0) ,\label{fren}
\end{eqnarray}
where $\epsilon = P_f-P_{\rm MC}\simeq n_{\rm cr}(\mu_{i}-\mu_f)$ is expressed through the (final)  value of the
pressure after the first-order phase transition has occurred and the pressure at the MC, $\mu_i$ and $\mu_f$ are the chemical potentials of the initial and final configurations (at fixed P and T ).
 The quantity  $\epsilon\neq 0$, if one deals with a first-order phase transition, and $\epsilon =0$, if a transition is of the second order. The maximum of the quantity
$\epsilon$ is ${\epsilon}_{m}= 4\lambda v^3 /(3\sqrt{3})$.
For the description of phase transitions to the uniform state, $k=0$, one may retain only the term $\propto c[\nabla (\delta n)] ^2$ in the expansion of the free energy in the density gradients using $c>0$.
For the description of phase transitions to the non-uniform state, $k_0\neq 0$, one should perform expansion  retaining at least terms up to   $\propto d[\Delta (\delta n)] ^2$ assuming $c<0$ and $d>0$. Therefore the last term in (\ref{fren}) appears only, if  $k=k_0 \neq 0$ \cite{Voskresensky:1993ux}, like for the phase transition to the solid state, liquid crystal state, or a pion condensate state in a dense nuclear matter.
Then for $k_0\neq 0$ and $c<0$, $d>0$ we have
\begin{eqnarray}
\delta F_{\rm L} (k_0)=\int \frac{d^3 x}{n_{cr}}\left\{ \frac{dm^*}{2} (\Delta \delta n)^2+\left(\frac{cm^* k_0^2}{2}- \frac{dm^* k_0^4}{2m^*}\right) (\delta n)^2 \right\}\,,
\end{eqnarray}
where $k_0^2=-\frac{c}{2d}>0$ follows from minimization of $\delta F_{\rm L} (k_0)$. In case of the phase transition to the uniform state one should put $k_0 =0$, $d=0$ (then $\delta F_{\rm L} (k_0)=0$)  and $c>0$. Then the first term $\propto c$ in Eq. (\ref{fren}) is associated with the positive surface tension, $\delta F_{\rm L}^{\rm surf}=\sigma S$, where $S$ is
the surface of the seed.

\begin{figure}
\centerline{%
\rotatebox{0}{\includegraphics[height=10.0truecm] {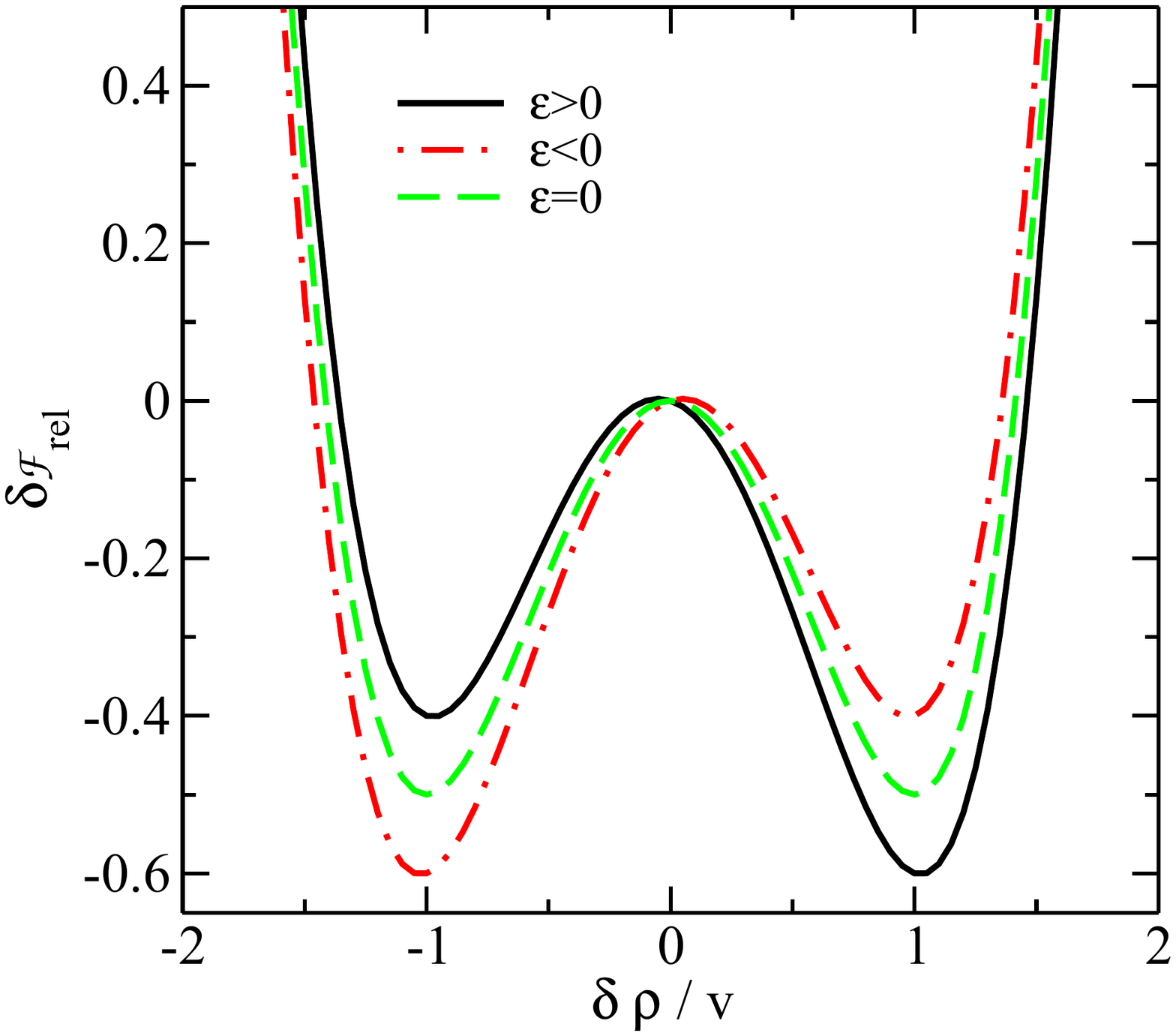}
} \rotatebox{0}{\includegraphics[height=10.0truecm]
{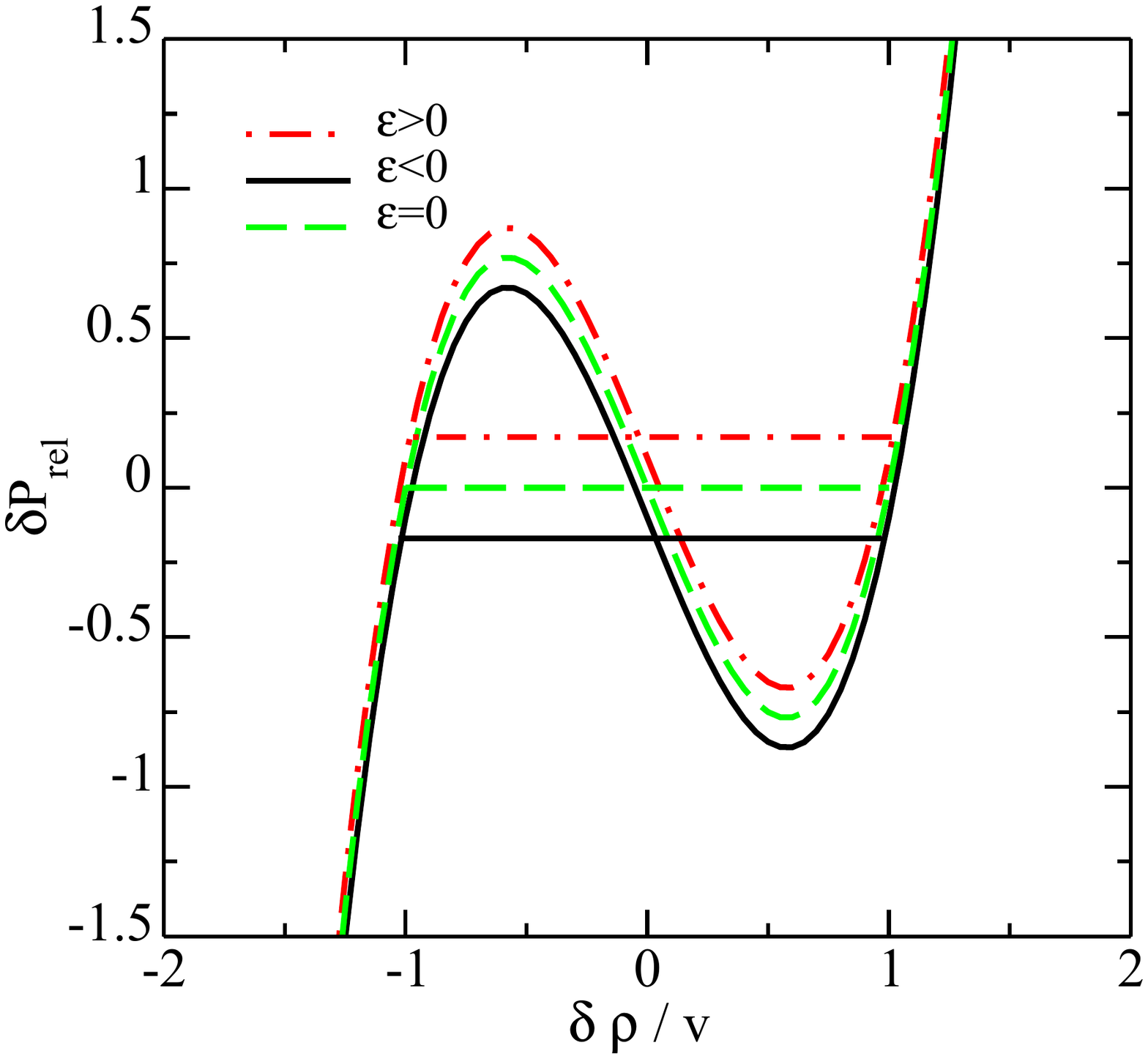}   }  } \caption{ The Landau free energy density
$\delta {\cal F}_{\rm rel} =  \delta {\cal{F}}_{\rm L} /{\cal{F}}_{\rm L} (T_{cr},\rho_{cr})$ and the value
$ \delta { P}_{\rm rel} = \rho_{cr}\frac{\delta [F_{\rm L} (T,\delta\rho )]}{\delta
(\delta\rho)}|_{T}/P (T_{cr},\rho_{cr})$, as functions of the order parameter $\delta
\rho =m^*\delta n$ for the EoS determined  by Eq. (\ref{fren}), at $T<T_{cr}$.
Dash horizontal line ($\epsilon =0$) in the right panel shows
MC. Figure is adopted from \cite{Skokov:2009yu}.}\label{pres}
\end{figure}

The Landau free energy density and pressure as functions of the order parameter $\delta
\rho$ for the equation of state determined  by Eq. (\ref{fren}) are shown in Fig. \ref{pres}. For $\epsilon =0$ two
minima of the Landau free energy coincide
and correspond to the MC on the curve $\delta P
(1/\rho)$ (shown by horizontal lines in the plot $\delta P
(\delta\rho)$ in the right panel). If  in the initial state  $(\delta\rho)_i = \rho_i -\rho_{cr}=0$, we deal with the spontaneous symmetry breaking and the second-order phase transition. For $(\delta \rho)_i =\rho_i -\rho_{cr}\neq 0$, $\epsilon
>0$ or $\epsilon <0$, we may deal either with the first-order phase transition from the metastable to the stable state , if $\rho_i$ corresponds to the metastable state, or with the second order phase transition either to the metastable state or to the stable state, otherwise.
For $\epsilon
>0$ (solid lines) the liquid state is stable and the gas state is metastable (SV),
and for $\epsilon <0$ (dash-dotted lines) the liquid state is
metastable (OL), whereas the gas state is stable. The dynamics of the transition starting from a point within spinodal region for $\epsilon \neq 0$ (but small) is described similarly to that for the second-order phase transition for $\epsilon =0$.

For the purely van der Waals equation of state (in this case $k_0=0$) one gets \cite{Skokov:2009yu}:
\begin{eqnarray}\label{parame}
v^2 (T) =- 4 \frac{\delta{{T}} n_{cr}^2 m^{*\,2}}{T_{cr}} ,
\quad \sigma =\sigma_0 \frac{|\delta
T|^{3/2}}{T_{cr}^{3/2}}\,,\quad  \sigma_0^2
 =32 m^{*}n_{cr}^2T_{cr} c.
\end{eqnarray}
Applying operator $\mbox{div}$ to  Eq. (\ref{Navier}) and replacing $\mbox{div}\vec{u}$ from Eq.~(\ref{contin}) for small $\delta \rho$ and $u$, keeping only linear terms in ${u}$, that is legitimate, since near the critical point processes develop slowly ($v^2 \propto -\delta T$),  we rewrite Eq.~(\ref{Navier}) as
 \begin{eqnarray}\label{v-t1}
-\frac{\partial^2 \delta n}{\partial t^2}&=\Delta\left[c\Delta
\delta n +\lambda v^2 \delta n -\lambda m^{*2} (\delta n)^3
+\epsilon/m^*  -(m^*n_{cr})^{-1}\left(\widetilde{\nu}{\eta_{cr}}
+{\zeta_{cr }} \right)\frac{\partial \delta n}{\partial
t}\right]\\
&-\Delta \left[ d\Delta^2 \delta n +(ck^2_0 +dk_0^4)\delta n \right] \,,\nonumber
 \end{eqnarray}
 $\widetilde{\nu}={2(\nu-1)}/{\nu}$, cf. \cite{Migdal:1990vm,Voskresensky:1993ux,Skokov:2009yu}.
Second line  in Eq. (\ref{v-t1}) yields  non-zero term only for the description of the condensation to the inhomogeneous state.

Consider $T<T_{cr}$. In  the dimensionless
variables $m^* \delta n =v \psi$, ${\tau}=t/t_0$, $\xi_i =x_i /l$, $i =1 ,\cdots ,
\nu$, $\nu =3$ for seeds of spherical geometry, Eq. (\ref{v-t1}) is presented as
 \begin{eqnarray}\label{dimens}
 &&- \beta \frac{\partial^2 \psi }{\partial
{\tau}^2} =\Delta_{\xi}\left(\Delta_{\xi}\psi +2\psi
 (1-\psi^2)+\widetilde{\epsilon}- \frac{\partial \psi}{\partial
 {\tau}}-  \frac{\lambda v^2 d}{2c^2}\Delta_{\xi}^2\psi +\frac{2(ck_0^2+dk_0^4)}{\lambda v^2}\psi   \right),\\
 &&l=\left(\frac{2c}{\lambda v^2}\right)^{1/2}\,,\,\,\,
  t_0
 =\frac{2\tilde{\eta}_{\rm r}}{\lambda v^2
 }\,,\,\,\,
 \widetilde{\epsilon}=\frac{2\epsilon}{\lambda v^3}\,, \,\,\, \beta
 =\frac{c }{\tilde{\eta}_{\rm r}^2 }\,,\,\,\,\tilde{\eta}_{\rm r}=\frac{(\tilde{\nu}\eta_{\rm r} +\zeta_{\rm r}) }{m^* n_{ cr}}.\nonumber
  \end{eqnarray}

It is important to notice that even for $k_0=0$,
Eq.~(\ref{dimens}) differs in the form from the standard Ginzburg-Landau  equation broadly exploited   in  the condensed matter physics, since Eq.~(\ref{dimens}) is of the second-order in time derivatives, whereas the standard Ginzburg-Landau  equation is of the first order in time derivatives.
The difference  disappears, if one puts   the  bracketed-term in
the r.h.s. of Eq.~(\ref{dimens}) to zero.  This procedure is however not legitimate at least for description  of the order parameter on an initial time-stage, since two initial conditions, such as $\delta n (t=0,\vec{r})=0$ and
$\partial_t \delta n (t,\vec{r})|_{t=0}\simeq 0$, should be fulfilled to describe an initially formed  fluctuation (seed). Thereby at least there exists an initial stage of the dynamics of seeds ($t\lsim t_{\rm init}$), which  is not described by the standard Ginzburg-Landau equation \cite{Skokov:2008zp,Skokov:2009yu}. The  bracketed-term in
the r.h.s. of Eq.~(\ref{dimens}) can indeed be put  zero, see below, if one considers   an effectively very viscous medium at   $\tau \gg 1$.
 Also note that Eq.~(\ref{dimens}) with the   bracketed-term in
the r.h.s. equal zero can be derived from the first-gradient order kinetic equation of Kadanoff-Baym \cite{Voskresensky:2010qu}.

Eq. (\ref{dimens}) should be supplemented by  Eq. (\ref{therm}) for the
heat transport, which owing to
Eq.~(\ref{contin})  after its linearization reads as
\begin{equation}\label{v-tS}
T_{cr}\left[
\partial_t \delta s
-s_{cr}(n_{cr})^{-1}
\partial_t \delta n
\right]=\kappa_{\rm r}\Delta \delta T\,.
\end{equation}
The variation of the temperature is related to the variation of
the entropy density $s[n,T]$ by 
\begin{equation}
\delta T \simeq T_{cr} (c_{V ,\rm r})^{-1}\left(\delta s
-({\partial s}/{\partial n})_{T, cr}\delta n\right)\,,
\end{equation}
$c_V$ is the density of the heat copacity.

\subsection{Typical time scales}
Let us perform some rough dimensional estimates of typical time scales in the problem. The evolution of a seed of one phase in another phase  is governed by  the slowest mode ($\delta\rho$ or $\delta s$, respectively).
The time scale  for the relaxation of the density following Eq.
(\ref{dimens}) is  $t_0\propto \tilde{\eta}$. Thus the non-zero viscosity plays the role of the driving force managing the time evolution of the density-mode. Also  $t_0\propto 1/(T_{cr}-T)$. Thereby the processes are slowed down near the critical point of the phase transition. The time scale  for the
relaxation of the entropy/temperature mode, following (\ref{v-tS}), is
 \begin{eqnarray}\label{kT}
t_T = R^2_{\rm seed} c_{V,\rm r} /\kappa_{\rm r} \propto R_{\rm seed}^2,
 \end{eqnarray}
 i.e., relaxation time of the temperature/entropy is proportional to the surface of the seed.
 Thus,
for $t_T (R_{\rm seed})<t_{0}$, i.e for  $R_{\rm seed}<R_{\rm fog}=\sqrt{\kappa_{\rm r}t_0/c_{V,\rm r}}$, where $R_{\rm fog}$
is  the typical size of the seed  at  $t\sim t_{0}=t_T$, dynamics of
seeds  is controlled by Eq. (\ref{dimens}) for the density mode.
For seeds with sizes $R_{\rm seed}>R_{\rm fog}$, the quantity $t_T\propto R_{\rm seed}^2$ exceeds
$t_{0}$ and growth of seeds is slowed down. Thereby,
the number of seeds with the typical size $R_{\rm seed}\sim R_{\rm fog}$ is increased with passage of
time, and a state of  fog is formed. For the quark-hadron phase transition in energetic heavy-ion collisions one \cite{Skokov:2009yu} estimates $R_{\rm
fog}\sim (0.1-1)$ fm and for the nuclear liquid-gas transition at low energies $R_{\rm fog}\sim
(1-10)$ fm $\lsim R (t_{\rm f.o.})$, where $R (t_{\rm f.o.})$ is the size of the fireball at the freeze out,
 $t_{\rm f.o.}$ is the fireball evolution time till freeze out.

There are only two dimensionless parameters in Eq.
(\ref{dimens}),  $\widetilde{\epsilon}$ and $\beta$.
The parameter $\widetilde{\epsilon}$ is responsible for a difference
 between   the Landau free energies of the  metastable  and  stable
states. For $t_{0}\gg t_T$ (isothermal stage), $\widetilde{\epsilon}\simeq const$
and  dependence on this quantity disappears because of
$\Delta_{\xi}\widetilde{\epsilon}\simeq 0$.
 Then, dynamics is controlled  by the  parameter
$\beta$,
 which characterizes an inertia. It is expressed in terms
of the surface tension and the viscosity as
 \begin{eqnarray}
\beta
 = (32T_{cr})^{-1}[\widetilde{\nu}\eta_{\rm r} +\zeta_{\rm r} ]^{-2}\sigma_0^2 m^{*}.\label{betasigma}
   \end{eqnarray}
The larger viscosity and the smaller surface tension, the
effectively more viscous (inertial)  is the fluidity of seeds.
For $\beta \ll 1$ one deals with the regime of effectively viscous (inertial)
fluidity and at $\beta \gg 1$ one deals  with the regime of almost perfect fluidity. Estimates \cite{Skokov:2009yu} show that for the nuclear liquid-gas phase transition typically $\beta\sim 0.01$.  For the
 quark-hadron   transition $\beta\sim 0.02-0.2$, even  for  very low
value of the ratio $\eta/s\simeq 1/(4\pi)$. The latter quantity characterizes fluidity of the matter at ultra-relativistic heavy-ion collisions \cite{Romatschke:2007mq}.  Thus, as we argued, in case of baryon-rich matter one deals with
effectively very viscous (inertial) evolution of density fluctuations  both
in cases of the nuclear liquid-gas and quark-hadron phase transitions.

In neutron stars an overcritical pion-condensate drop reaches a size $R\sim 0.1$ km for $t\sim 10^{-3}$ sec. by the growth of the density mode. Then it may reach $R\sim (1-10)$ km for typical time $t_T$ varying from $\sim 10$ sec.  up to several hours (rather than for typical collapse time $\sim 10^{-3}$ sec). A delay appears owing to neutrino heat transport to the surface
(effect of thermal conductivity) that strongly depends on the value of the pion softening, which is stronger for most massive neutron stars \cite{Migdal:1990vm}. One should also take into account that that the bulk viscosity is significantly increased in presence of soft modes \cite{ML37,LL06}, e.g., near the pion condensation critical point \cite{Kolomeitsev:2014gfa}.
Also notice that description of the dynamics of the pion-condensate phase transition  is specific, since the transition occurs to the inhomogeneous  liquid-crystal-like state characterized by  $\vec{k}\neq 0$. Seeds of  the liquid-crystal-like state prove to be elongated in the process of their growth \cite{Voskresensky:1993ux}, similar  effect is observed in liquid crystals.

Thus,
interplay between viscosity, surface tension, and thermal
conductivity effects is responsible for the typical time and size scales of
fluctuations.

\subsection{Stationary solutions}

Now let us find stationary solutions of Eq. (\ref{v-t1}). For the condensation in the state $k\neq 0$ we find solution in the form
\begin{eqnarray}
m^* \delta n=a[\sin (k x+\chi) +\frac{c_1 \widetilde{\omega}^2 (k^2)}{\widetilde{\omega}^2 (9k^2)}\sin (3k x +\chi) +...]+O(\epsilon)\,,\label{sinExp}
\end{eqnarray}
where $\chi$ is a constant phase,
\begin{eqnarray}
\widetilde{\omega}^2 (k^2)=-\lambda v^2 +ck^2 +dk^4-ck_0^2 -dk_0^4\,.\label{omtilde}
\end{eqnarray}

For the condensation in the uniform state $k_0\neq 0$, $c<0$, $d>0$ the gap $\widetilde{\omega}^2 (k^2)$ has a minimum for $k=k_0$.
The phase transition arises for $\widetilde{\omega}^2 (k_0^2)<0$.
Setting (\ref{sinExp}) in Eq. (\ref{v-t1}) we find
\begin{eqnarray}
a^2 =-\frac{4}{3} \widetilde{\omega}^2 (k_0^2)/\lambda>0, \quad c_1 =-1/3\,.
\end{eqnarray}
Minimization of the free energy in $k$ yields $k=k_0$ and $\widetilde{\omega}^2 (k_0^2)=-\lambda v^2$,
$\widetilde{\omega}^2 (k_0^2)>0$ for $T>T_{cr}$ and $\widetilde{\omega}^2 (k_0^2)<0$ for $T<T_{cr}$, $\widetilde{\omega}^2 (9k_0^2)=-\lambda v^2 +16 c^2/d\gg |\widetilde{\omega}^2 (k_0^2)|$. Thereby with appropriate accuracy we may use $\delta n\simeq a[\sin (k_0 x+\chi)$ that yields $\delta F_L (k_0)\simeq -\lambda v^4 V/(6m^*n)+O(\epsilon^2)$, where V is the volume of the system. Thus solution (\ref{sinExp}) describes the stationary state at the second-order phase transition.

For the condensation in the uniform state $k_0=0$ we have \cite{Skokov:2009yu}
\begin{eqnarray}
\widetilde{\omega}^2 (k^2)=-\lambda v^2 +ck^2\,,\quad  k^2<\lambda v^2/c\,,\quad c>0\,.
\end{eqnarray}
 Two spatially constant stationary solutions minimizing the free energy for $T<T_{cr}$ correspond to $k=0$. They describe metastable and stable states:
\begin{eqnarray}
\delta n_{\rm st}\simeq \pm v/m^* +\epsilon/(2\lambda v^2 m^*)\,.\label{solConst}
\end{eqnarray}
The free energy corresponding to these solutions is given by
\begin{eqnarray}
\delta F_{\rm L} (k=0, k_0=0)\simeq -\frac{\lambda v^4 V}{4m^*n_{cr}} \left(1\pm \frac{4\epsilon}{\lambda v^3}\right)\,.\label{FLknul}
\end{eqnarray}
For $k\neq 0$ solutions in the form (\ref{sinExp}) are valid  for $|\widetilde{\omega}^2 (k^2)|\ll \widetilde{\omega}^2 (9k^2)$ and they yield for $k_0=0$:
\begin{eqnarray}
  \delta F_{\rm L} (k\neq 0, k_0=0)\simeq -\frac{\lambda v^4 (1-ck^2/(\lambda v^2))V}{6m^*n_{cr}}\,.\label{FLknulk}
  \end{eqnarray}
 Although the minimum of the free energy for $k_0=0$  is given by (\ref{FLknul}) corresponding to  solutions (\ref{solConst}) obtained for  $k=0$,  rather than by solutions of (\ref{sinExp}) corresponding to the free energy  (\ref{FLknulk}), nevertheless, as we will demonstrate below, solutions (\ref{sinExp}) characterising by $k\neq 0$ have a physical meaning.

\section{Dynamics of seeds at first-order phase transition from metastable state to stable state}\label{metastable}
Consider the limit of a high thermal conductivity, when in Eq. (\ref{v-t1}) the temperature can be put constant. Solution of Eq. (\ref{v-t1}) describing
dynamics of the density  fluctuation  developing from the metastable state to the stable state is then presented in the form \cite{Skokov:2010dd}
 \begin{equation}\label{delr} \delta n (t,r)\simeq
\frac{v(T)}{m}\left[\pm\mbox{th} \frac{r-R_{\rm seed}
(t)}{l}+\frac{{\epsilon}}{2\lambda v^3(T)}\right]+(\delta
n)_{\rm cor},
\end{equation}
where the upper sign corresponds to the evolution  of  bubbles of the gas
and the lower-sign solution describes  evolution of  droplets of liquid for $\nu =3$, and the
solution is valid for $|\epsilon/(\lambda v^3(T))|\ll 1$.
 Compensating correction  $(\delta
n)_{\rm cor}$ is introduced to  fulfill the   baryon
number conservation. Considering spatial coordinate $r$ in the vicinity of a
bubble/droplet boundary we get  equation describing evolution of the
seed size \cite{Skokov:2009yu,Skokov:2010dd}:
\begin{equation}\label{dim}
\frac{m^{*2}\beta t_0^2}{2l}
\frac{d^2R_{\rm seed}}{dt^2}=m^{*2}\left[\frac{3\epsilon}{2\lambda v^3 (T)
}-\frac{2l}{R_{\rm seed}}\right]-\frac{m^{*2} t_0}{l}\frac{d R_{\rm seed}}{dt}.
\end{equation}
This equation reminds the Newton second law for a one-dimensional system, where the quantity $M=\frac{m^{*2}\beta t_0^2}{2l}\propto {(T_{cr}-T)^{-3/2}}$ has a meaning of a mass, $m^{*2}[\frac{3\epsilon}{2\lambda v^3 (T)
}-\frac{2l}{R_{\rm seed}}]$ is an external force and $-\frac{m^{*2}t_0}{l}\frac{d R_{\rm seed}}{dt}$ is the friction force, with a viscous-friction coefficient  proportional to an effective viscosity and inversely proportional to $\sqrt{T_{cr}-T}$. Following Eq. (\ref{dim})
 a bubble of an overcritical  size $R_{\rm seed}>R_{cr}=4l\lambda v^3(T)/(3\epsilon)$
 of the stable gas
phase, or respectively a droplet of the stable liquid  phase, been initially
prepared in a fluctuation inside a metastable phase,  then grow. On an early stage of the evolution  the size of
the overcritical bubble/droplet  $R_{\rm seed}(t)$ (for $R_{\rm seed}>R_{cr}$) grows with an acceleration.
Then it reaches a steady growth regime with a constant velocity
$u_{\rm as}=\frac{3\epsilon l}{\lambda v^3 (T) t_0}\propto |(T_{cr} -T)/T_{cr}|^{1/2}$.
 In the interior of the
seed $\delta n \simeq \mp v(T)/m^*$. The
correction $(\delta n)_{\rm cor} \simeq v(T)R_{\rm seed}^3(t)/ (m^* R^3)$
is very small for $R_{\rm seed}(t)\ll R$, where $R$ is the radius of the whole system. In cases of the quark-hadron and nuclear liquid-gas phase transitions in heavy-ion collisions  $R(t)$ is the radius of the expanding fireball. Usage of the isothermal approximation in Eq.~(\ref{delr}) needs fulfillment of inequality $t_{\rho}\sim \frac{R_{\rm seed}(t_{\rm f.o.}) }{u_{\rm as}}\gg t_T$. For $R_{\rm seed}\sim R_{cr}$ and for $\epsilon \sim \epsilon_m$ we get $t_{\rho}\sim t_0$, and isothermal approximation is valid for $R_{\rm seed}<R_{\rm fog}$. For $\epsilon \ll \epsilon_m$ we get $t_{\rho}\gg t_0$ and isothermal approximation remains correct for seeds of the size $R_{\rm seed}<R_{\rm fog}\epsilon_m/\epsilon$.

Substituting Eq.~(\ref{delr}) to Eq.~(\ref{v-tS}) for $T\simeq$ const (that is correct in linear approximation) we obtain
 \begin{eqnarray} \label{deltasinf}
&&\delta s = \left(\frac{\partial s}{\partial
n}\right)_{T}\left\{\frac{v(T)}{m}\left[\pm\mbox{th} \frac{r-R_{\rm seed}
(t)}{l}+\frac{{\epsilon}}{2\lambda_{cr} v^3 (T)}\right]+(\delta
n)_{\rm cor}\right\}.
 \end{eqnarray}

Note that for the description of expanding fireball formed in heavy-ion collisions  the approximation of a quasi-adiabatic expansion  can be used  even in presence of the weak first-order phase transition (for $\delta s \ll s$ and $\delta n \ll n$). The evolution of droplets/bubbles in metastable region can be considered at fixed size of the fireball  provided  $t_{\rm f.o.}\gg (t_\rho, t_T)$.

\section{Dynamics of fluctuations in unstable region}\label{unstable}
\subsection{Growth of fluctuations of small amplitude. Linear regime}

In this section the ``r''-reference point can be taken
arbitrary, therefore we suppress the subscript   ``r''.
To find solutions of
the linearized hydrodynamical equations we put, cf. \cite{Skokov:2010dd},
\begin{eqnarray}\label{delns}
\delta n =\delta n_0 \mbox{exp}[\gamma t +i \vec{k}\vec{r}] -\frac{\epsilon}{m^*\lambda v^2}\,,,
\quad \delta s= \delta s_0 \mbox{exp}[\gamma t +i
\vec{k}\vec{r}],
\quad T =T_{>}+\delta T_0
 \mbox{exp}[\gamma t +i \vec{k}\vec{r}],
\end{eqnarray}
where $T_{>}$ is the temperature of the uniform matter. For $|\delta n|\gg |\frac{\epsilon}{m^*\lambda v^2}|$, i.e. for $\epsilon \ll \epsilon_m$, description of a fluctuation  in spinodal region at the first-order phase transition and for description  of the second-order phase transition are the same and we may put $\epsilon \to 0$.
 Then from
linearized equations of non-ideal hydrodynamics (\ref{v-t1}),  (\ref{v-tS}) we find
 the increment,
$\gamma (k)$,
 \begin{eqnarray}\label{incr} &&\gamma^2 =
-k^2 \left[\widetilde{\omega}^2 (k^2) +\tilde{\eta}\gamma
+\frac{u_{\tilde{s}}^2 -u_T^2 }{1+\kappa k^2
/(c_{V}\gamma)}\right], \end{eqnarray}
where $\tilde{\eta}=\frac{(\tilde{\nu}\eta +\zeta) }{m^* n}$.
Eq. (\ref{incr}) has three solutions corresponding to the growth of the density and thermal modes.
For $\kappa k^2
/(c_{V}|\gamma|)\gg 1$   the temperature in the seed can be put constant and we may deal with only one equation for the density mode (\ref{v-t1}) that yields
\begin{eqnarray}\label{incrn} &&\gamma^2 =
-k^2 \left[\widetilde{\omega}^2 (k^2) +\tilde{\eta}\gamma
\right]\,, \end{eqnarray}
from where we find two  solutions for the density-modes,
\begin{eqnarray}\label{Solincr}
\gamma_{1,2} =-\frac{k^2\tilde{\eta}}{2}\pm \sqrt{\frac{k^4\tilde{\eta}^2}{4}-k^2 \tilde{\omega}^2 (k^2)}\,.
\end{eqnarray}
For $\tilde{\omega^2 (k^2)}<0$, that corresponds to  the region of the phase transition,  the upper-sign solution, $\gamma_1 >0$,  describes the growing mode and the lower sign solution, $\gamma_2 <0$, describes the damping mode.
For $k^2 \tilde{\eta}^2/|\tilde{\omega}^2 (k^2)|\ll 1$ we have
\begin{eqnarray}\label{gam1lowvisk}
\gamma_1 \simeq \sqrt{-k^2\tilde{\omega}^2 (k^2)}-\frac{k^2\tilde{\eta}}{2}+O(k^3\tilde{\eta}^2/|\tilde{\omega} (k^2)|)
 \end{eqnarray}
 for the growing mode.  In the opposite limit $k^2 \tilde{\eta}^2/|\tilde{\omega}^2 (k^2)|\gg 1$ we obtain
\begin{eqnarray}\label{gam1highvisk}
\gamma_1 \simeq -\tilde{\omega}^2 (k^2)/\tilde{\eta}+O(\tilde{\omega}^4 (k^2)/(k^2\tilde{\eta}^3))\,.
\end{eqnarray}
Note that in condensed matter physics  a transition from a liquid to a glass state can be interpreted as a first-order phase transition occurring within a  spinodal region at a very high viscosity  \cite{Voskresensky:1993ux}. Then there appears an order at several \AA - scale, which transforms in a disorder at larger distances.

For  $k_0=0, c>0$, for the most rapidly growing mode (for $\gamma_m =\mbox{max}\{\gamma_1\}$ corresponding to $k=k_m$)  we find
  \begin{eqnarray}\label{gam12m}
 \gamma_{m} \simeq \frac{\lambda v^2  }{(2\sqrt{\beta}+1)\tilde{\eta}}\,, \quad k_m^2 \simeq \frac{\lambda v^2\sqrt{\beta}}{(2\sqrt{\beta}+1)c}\,.\nonumber
 \end{eqnarray}

 For $k_0\neq 0, c<0, d>0$ the most rapidly growing mode corresponds to $k =k_0$, then $\tilde{\omega}^2 (k_0^2)<0$ and $|\tilde{\omega}^2 (k_0^2)|$ as a function of $k^2$ is the largest.

\subsection{Growth of fluctuations of arbitrary amplitude. Nonlinear regime}
Now we will find solution of the non-linear Eq.  (\ref{v-t1}).
We search the solution
in the form
\begin{eqnarray}
m^* \delta n=a f(t) \left[\sin (k x+\chi) +\frac{c_1 \widetilde{\omega}^2 (k^2)}{\widetilde{\omega}^2 (9k^2)}\sin (3k x +\chi) +...\right]+O(\epsilon)\,,\label{sinExpnonlin}
\end{eqnarray}
as (\ref{sinExp}) with $a^2 =-\frac{4\widetilde{\omega}^2}{3\lambda}>0$ but now with  $f(t)$ satisfying equation
\begin{eqnarray}\label{ft}
\partial_t^2 f =-k^2 \widetilde{\omega}^2 (k^2) f(1-f^2) -k^2 \tilde{\eta} \partial_t f\,.
\end{eqnarray}
For $k^2 \tilde{\eta}^2/|\tilde{\omega}^2 (k^2)|\gg 1$, i.e. for $\beta \ll 1$ or $\tilde{\eta}\gg \sqrt{c}$,   the term  $\partial_t^2 f$ in the l.h.s. of Eq. (\ref{ft}) can be dropped and the amplitude
\begin{eqnarray}\label{finterp}
 f(t)=\frac{f_0 e^{\gamma t}}{\sqrt{1+f_0^2 e^{2\gamma t}}}\,,
\end{eqnarray}
 fulfils  the resulting  Eq. (\ref{ft}),  $f_0/\sqrt{1+f_0^2}$ shows the amplitude of the  fluctuation at $t=0$,  $f_0$ is an arbitrary constant. For  $k\sim k_m$ at $k_0=0$ this solution  holds for $k_m^2 \tilde{\eta}^2/|\tilde{\omega}^2 (k_m^2)|\gg 1$. For $k=k_0\neq 0$ the criterion of applicability renders as $k_0^2 \tilde{\eta}^2/|\tilde{\omega}^2 (k_0^2)|\gg 1$. In both cases $k_0=0$ and $k_0\neq 0$ with the density distribution given by  (\ref{sinExpnonlin}), (\ref{finterp})
 the free energy renders
\begin{eqnarray}\label{freeent}
\delta F_{\rm L} (t) =-\frac{V\widetilde{\omega}^4(k^2)}{6\lambda m^* n}f^2(t)\left(2-f^2(t)\right)\,.
\end{eqnarray}
For $t\to \infty$ we have $f(t\to\infty)\to 1$ and $\delta F_{\rm L}$ reaches the minimum. For $k=k_0$ this value coincides with (\ref{FLknulk}) given by the stationary solution.

In general case expression (\ref{finterp}) yields an interpolation between  two approximate  solutions valid for the limit cases  $\gamma t\ll 1$ and  $\gamma t \gg 1$. Replacing (\ref{finterp}) in Eq. (\ref{ft}) we obtain then the same solutions (\ref{Solincr}) as in linear case.

Let first $k_0 =0$. For $t\to \infty$ using solution (\ref{finterp}) at $\gamma =\gamma_m =\gamma(k_m)$ we find
\begin{eqnarray}\label{FLgamm}
\delta F_{\rm L} (t\to \infty)=-\frac{\widetilde{\omega}^4 (k_m^2)V}{6\lambda m^* n}.
\end{eqnarray}
 For the case  of a large effective viscosity/inertia, $\beta \ll 1$, we obtain $\delta F_{\rm L} (t\to \infty)\simeq -\frac{\lambda v^4 V}{6 m^* n}$ that coincides with (\ref{FLknulk}) but is still larger than the value given by (\ref{FLknul}). For the case   of a  small  effective viscosity/inertia, $\beta \gg 1$, we find $\delta F_L (t\to \infty)\simeq -\frac{\lambda v^4 V}{24 m^* n}$ that is much higher than the free energy given by both stationary solutions (\ref{FLknul}), (\ref{FLknulk}).  Thus one may expect that expression (\ref{FLgamm}) either describes  a metastable state or a state, which slowly varies on a time  scale $t_k \gg t_\gamma \sim 1/\gamma_m$, reaching for $t\gg t_k$ the stationary state with the free energy given by (\ref{FLknul}). To show the latter possibility consider the case $\beta \gg 1$ and assume $k$ in solution (\ref{sinExpnonlin}) to be a slow function of time, i.e. $k=k(t)$, for typical time scale $t_k\gg t_\gamma$.
One can see that for   $R_{\rm seed}\ll t_k |u_T|$  the quantity  $k(t)$ satisfies equation
$(d^2k/dt^2) =-k^2 \tilde{\eta}(dk/dt)$ with the solution
\begin{eqnarray}\label{kt}
k(t)=k_{00}[1+\tilde{\eta}\lambda v^2 t/(3c)]^{-1/2}
\end{eqnarray}
such as $k(t\to \infty)\to 0$ and the free energy for $t\to \infty$ indeed reaches the limit (\ref{FLknul}) provided we set $\sin \chi \simeq \frac{\sqrt{3}}{2} -\frac{\sqrt{3}\,m^*\tilde{\epsilon}}{8}$.
From Eq. (\ref{kt}) we easily find that the typical time scale is
 $t_k \sim \beta t_0$ and we check that indeed $t_k\gg t_\gamma$. For $R_{\rm seed}\gsim t_k |u_T|\sim l\sqrt{\beta}$ the solution (\ref{sinExpnonlin}) with (\ref{kt}) does not hold and should be modified.

 For $\beta \ll 1$, $k_0=0$,  expression (\ref{kt}) with slowly varying $k(t)$ does not hold.  At  realistic conditions  convection and sticking processes (at sizes $\sim l$) may be allowed, which destroy periodicity, and owing to these processes the system may finally reach the ground  state with the free energy given by  (\ref{FLknul}). Thus one possibility is that for the typical  time $t\sim t_\gamma\sim t_0$ the  quasiperiodic solution (\ref{FLgamm}) is formed with typical $k\simeq k_m$, corresponding to a metastable state with the free energy given by (\ref{FLknulk}). Such a distribution is formed most rapidly.    Another possibility is that for the  typical time scale $t_{unif}> t_\gamma$ in the system of a large size an approximately  uniform solution (\ref{unif}) is developed.
 In the latter case to proceed consider the case $k\sim 1/R\ll k_m$, where $R$ is the typical size of the system ($R=R_{f.o.}$ for
 the fireball formed in heavy-ion collisions). The spatially uniform solution of equation
 $$\Delta_\xi \psi+2\psi (1-\psi^2)+\tilde{\epsilon}=\partial_\tau \psi$$
  that follows from (\ref{dimens}) in this case (as well as for seeds of a  size $R_{\rm seed}\ll R$ at  $\beta \ll 1$, as we have argued above), is given by
 \begin{eqnarray}\label{unif}
 \psi (t) =\pm 1/\sqrt{1+e^{-\tau}{(1-\psi_0^2)}/{\psi_0^2}}\,,\end{eqnarray}
 where we for simplicity put $\tilde{\epsilon} \to 0$. Typical time needed for the initial amplitude $\psi_0\ll 1$ to grow to $\psi(t\to \infty)\simeq \pm 1$ is  $t_{unif}\sim t_0 \ln (1/\psi_0^2)\gg t_\gamma$.

 Thus,  we found some novel solutions describing evolution of fluctuations in the region of instability
 additionally  to the uniform solution (\ref{unif}). For $k=$const$\neq 0$ we found  periodic solutions  given by (\ref{sinExpnonlin}), (\ref{finterp}). For $k=k_0\neq 0$ the solution yields minimum of the free energy for $t\to \infty$. For $k_0=0$, $\beta \gg 1$,  we found quasiperiodic  solutions (\ref{sinExpnonlin}), (\ref{finterp}) with    $k=k(t)$ from (\ref{kt}), yielding minimum of the free energy for $t\to \infty$.

\section{Conclusions}\label{Conclusion}

According to  our findings,
signatures of QCD spinodal instabilities may be observed
in
experiments with heavy ions in some collision energy interval that corresponds
to the first-order phase transition region of the QCD phase diagram. If  typical time of the growth of a fluctuation in unstable region $t_\gamma$ and of the fireball expansion $t_{\rm f.o.}$ satisfy inequality $t_\gamma \lsim t_{\rm f.o.}$, one of the possible experimental signatures of the  spinodal region would be manifestation of a spatially quasiperiodic structure with a typical period $r\simeq 2\pi/k_m$  in the rapidity spectra.
If the parameter characterizing effective viscosity/inertia $\beta$ were $\gg 1$, cf. Eq. (\ref{betasigma}), for $t_\gamma \ll t_{\rm f.o.}$ one of the possible experimental signatures of the  spinodal region would be manifestation of spatially quasiperiodic fluctuations with a typical size $r\sim 2\pi/k (t_{\rm f.o.})$.
However rough estimates  done for the quark-hadron  and nuclear gas-liquid first-order phase transitions in heavy-ion collisions \cite{Skokov:2008zp,Skokov:2009yu,Skokov:2010dd} indicate that $\beta \ll 1$.

Concluding, we note that viscosity and thermal conductivity are driving forces of  the first-order liquid-gas and quark-hadron phase transitions to the state with $k_0=0$, and the spinodal instability occurs for $T$ below ITS line. The manifestation of a spatially quasiperiodic structures with a typical period $2\pi/k_m$, cf. Eq.  (\ref{sinExpnonlin}),  in the rapidity spectra in heavy-ion collisions in some collision energy interval  could be interpreted as a signature of the occurrence  of the spinodal instability at the first-order phase transition.
For the second-order phase transition to the state with $k_0\neq 0$, as for the case of the pion condensation in dense nuclear matter, the periodic solution (\ref{sinExpnonlin}) holds for $k=k_0\neq 0$, with $k_0$ not depending on time.


\reftitle{References}

\end{document}